\documentclass[aps,prb,preprint,groupedaddress]{revtex4}
\usepackage{graphicx}
\begin{document}
\title{Quantum Entanglement and Electron Correlation in Molecular Systems} 
\author{Hefeng Wang and  Sabre Kais \footnote{
Corresponding author:  kais@purdue.edu}}
\affiliation{Department of Chemistry, Purdue University, West Lafayette, IN 47907}
\begin{abstract}
\noindent

We study the relation between quantum entanglement and electron 
correlation in quantum chemistry calculations. 
We prove that the Hartree-Fock (HF) wave function 
does not  violate Bell's inequality, thus is not entangled
 while the configuration interaction (CI) wave 
function is entangled since it violates Bell's inequality.
Entanglement is related to electron correlation and 
might be used as an alternative measure of the electron correlation
in quantum chemistry calculations. As an example we show the calculations of 
entanglement for  the H$_2$ molecule and how it is related to  
electron correlation of the system, which is the difference between the exact and
the HF energies.

\end{abstract}
\maketitle
\clearpage

\section{Introduction}

Entanglement is a quantum mechanical property that describes the nonlocal correlation 
between quantum systems which has no classical analog\cite{EPR,diviccezo,entg1,Nielsen,gruska}.
It is a distinguishing feature of quantum mechanics and is regarded as a resource for many 
applications such as quantum communication and quantum computation. The basic criterion 
is that an entangled 
state violates Bell's inequality. A thorough understanding 
of entanglement has not been reached. The characterization 
of entanglement in a system of identical particles has proven particularly challenging. 
A number 
of entanglement measures for system of identical fermions have been 
proposed\cite{bar1,bar2,eck,zan,shi1,shi2} with focus
on different aspects of the phenomenon\cite{KAIS}.

Electron correlation in a chemical  
system is over all electrons and is nonlocal because 
of the identical particle property of electrons. 
Electron correlation in quantum chemistry is characterized in terms of 
the correlation energy, which is defined as the energy difference between the 
exact non-relativistic energy of the system and the Hartree-Fock (HF) energy obtained 
in the limit that the basis set approaches completeness\cite{lowdin}. 
The correlation energy is 
resulted from the Coulomb repulsion and Pauli's exclusion principle. In the HF theory, 
the ground state is assumed to be described by only one Slater determinant. 
In dealing with the 
electron correlation effect, the configuration interaction (CI) approach postulates that 
the total wave function of the system is a linear combination of Slater determinants. 
The correlation energy is calculated as the difference between the full CI (FCI) energy and
the HF energy using the same one-electron basis set, this is 
called the basis set correlation energy. 
As the one-electron basis goes to completeness, this basis set correlation 
energy approaches the exact correlation energy.

Vedral has pointed out that interaction in general gives rise to 
entanglement\cite{vedral}. Here,  we analyze the relation 
between the entanglement and correlation energy in molecular system. We show that the 
HF wave function, a single Slater determinant, does not violate  
Bell's inequality, thus it is unentangled. But the CI wave function violates the 
Bell's inequality, it is entangled. According to this we suggest 
that electron correlation is related to quantum entanglement. 
The result of the ab initio calculations for H$_2$ molecule 
is shown to support our suggestion. Since there is no operator 
in quantum chemistry for electron 
correlation energy, and measurement for quantum entanglement has been suggested, thus
entanglement might be used as an alternative way to measure the electron correlation.  

\section{Electron correlation in molecular systems} 

Electron correlation has a strong influence on many atomic, molecular\cite{wilson}, 
and solid properties\cite{march} of physical systems.  
For many-particle systems, like molecular systems, the state 
wave functions are vectors in 
Hilbert space of the system. Physically, these wave functions 
describe, under the Born-Oppenheimer approximation\cite{cohen} and 
the orbital approximation,  how the electrons are distributed between 
orbitals (one-electron wave functions).
A distribution of electrons between the orbitals,
obeying the Pauli exclusion principle, determines an electron 
configuration. Each configuration creates a set of configuration state functions (CSFs).
CSFs are constructed from a set of Slater determinants, 
the CSFs span the Hilbert space of the system.

The HF method\cite{rotn,szabo} assumes that the ground state wave 
function is a single Slater
determinant. But for most systems this is not sufficient, 
electron correlation needs to be considered in order to accurately describe 
a molecular system.
In quantum chemistry, electron correlation can be divided into two parts:
dynamic electron correlation and 
non-dynamic electron correlation.

If a number of electron configurations
are relatively close in energy (degenerate or near-degenerate), the HF 
approximation can not provide an adequate description of the ground state wave function.
This becomes apparent when one explores regions of Born-Oppenheimer potential energy
surfaces(PESs) far away from the equilibrium structure.
An accurate description should consider the expansion of all these configurations. A
description of the ground state using these degenerate or near-degenerate 
electron configurations comprises the non-dynamic electron correlation.
In quantum chemistry the multiconfigurational self-consistent field (MCSCF) 
method is used to consider the non-dynamic electron correlation. 
In this method both the molecular orbitals and the configuration expansion 
coefficients in a small configuration space are optimized variationally.

Electrons in a system tend to avoid each other because of the Coulomb repulsion,
this is called dynamic electron correlation. Configurations created by excitations
of electrons from the non-dynamically-correlated wave function can 
be considered with which to 
describe the dynamic correlation. In quantum chemistry, the CI 
approach is used to describe the dynamic electron correlation. If the wave 
function is expanded 
over all CSFs that constructed from all possible electron configurations, 
then this wave function is 
called the FCI wave function. In the FCI approach, all the configuration 
coefficients are optimized variationally.

For a given basis set, the most accurate wave function that considers both 
non-dynamic and dynamic electron correlation is a wave function in the FCI space 
In this wave function, each electron configuration will have
a certain contribution to the total wave function. This total wave function is:

\begin{equation}
\psi_{total}=\sum_{K}A_{K}\Phi_{K},
\qquad \Phi_{K}=A{\prod_{i\subset K}\phi_{i}},
\qquad \phi_i=\sum_{\mu}\chi_{\mu}C_{\mu i}
\end{equation}

which is a linear combination of all the CSFs in the full CI space. Each CSF differs in
how the electrons are distributed in the MOs, $\phi_i$. 
The MOs are expanded in a basis of atomic orbitals, $\chi_\mu$. 
For this wave function both the configuration expansion coefficients $A_K$ and
the MO expansion coefficients $C_{\mu i}$ are variationally optimized.
The optimized vector is the best approximation to the exact 
wave function of the system for a given one-electron basis set.

After the FCI wave function is obtained, 
the optimized result can be expressed in terms of the reduced one particle 
density matrix, which gives the occupation number on each orbital after 
considering electron correlation. Thus the reduced one particle density matrix contains
all the information about electron correlation.

\section{Entanglement in systems of identical particles}

Entanglement in a system of identical particles is
fundamentally different from that of
the distinguishable particles, for which entanglement is
invariant under local unitary transformations. In contrast,
there is no local operation that only acts on one of the identical particles\cite{shi1}.
In the identical particle system, the single-particle basis
transformation is made on each particle and a different set of 
particles is chosen in representing
the many-particle system. Thus the entanglement property of a system 
of identical particles depends on
the single-particle basis used\cite{shi1,vedral}.

In a system of fixed number of electrons, characterization of the 
entanglement must exclude the
non-factorization of the wave function due to antisymmetrization. It can be proved that
a wave function formed through antisymmetrization of a factorized product state, 
a Slater determinant,
does not violate Bell's inequality\cite{grd}, thus it is unentangled; while a CI wave 
function violates Bell's inequality, it is entangled.

For an $N$-electron system, the Slater determinant is:

\begin{eqnarray}
|\Psi(1,2,...,N)> & = & |\chi_{j}(1)\chi_{k}(2)...\chi_{l}(N)> \nonumber \\
                  & = & \frac{1}{\sqrt{N!}}
\left |\begin{array}{cc}
\chi_{j}(1) &\chi_{k}(1) ... \chi_{l}(1)\\
\chi_{j}(2) &\chi_{k}(2) ... \chi_{l}(2)\\
\vdots\\
\chi_{j}(N) &\chi_{k}(N) ... \chi_{l}(N)\\
\end{array}\right | \nonumber \\
\end{eqnarray}

where $\chi_j$s are the spin orbitals. We define an operator:

\begin{eqnarray} 
 \hat{O} & = & \vec{\sigma}(1) \vec{a} P_{j}(1){\otimes}
 \vec{\sigma}(2) \vec{b} P_{k}(2)+
 \vec{\sigma}(1) \vec{a} P_{j}(1)\otimes
 \vec{\sigma}(3) \vec{b} P_{k}(3)+...+\vec{\sigma}(1) \vec{a} P_{j}(1)\otimes
 \vec{\sigma}(2) \vec{b} P_{k}(N) + \nonumber \\
 &   & \vec{\sigma}(2) \vec{a} P_{j}(2)\otimes
 \vec{\sigma}(1) \vec{b} P_{k}(1)+
 \vec{\sigma}(2) \vec{a} P_{j}(2)\otimes
 \vec{\sigma}(3) \vec{b} P_{k}(3)+...+\vec{\sigma}(2) \vec{a} P_{j}(2)\otimes
 \vec{\sigma}(2) \vec{b} P_{k}(N) \nonumber \\
 &   & +...+ \nonumber \\
 &   & \vec{\sigma}(N) \vec{a} P_{j}(N)\otimes
 \vec{\sigma}(1) \vec{b} P_{k}(1)+
 \vec{\sigma}(N) \vec{a} P_{j}(N)\otimes
 \vec{\sigma}(2) \vec{b} P_{k}(2)+...+ \nonumber \\
 &   & \vec{\sigma}(N) \vec{a} P_{j}(N)\otimes
 \vec{\sigma}(N-1) \vec{b} P_{k}(N-1)
\end{eqnarray}

where $P$ is the projection operator, $\vec{a}$ and $\vec{b}$ are any real three-dimensional unit vectors,
the observable $\vec{\sigma}\vec{a}$ can be refered to as a measurement of spin along the $\vec{a}$ axis, it
gives the result $+1$ or $-1$.
We can calculate the expectation value $E(\vec{a},\vec{b})$:

\begin{eqnarray}
E(\vec{a},\vec{b}) & = & <\Psi(1,2,...,N)|\hat{O}|\Psi(1,2,...,N)> \nonumber \\
                   & = & <\chi_{j}|\vec{\sigma} \vec{a}|\chi_{j}><\chi_{k}|\vec{\sigma} \vec{b}|\chi_{k}>
\end{eqnarray}

in which it can be factorized as tensor product. The occurrence of a factorized product of two mean values
implies that no choice of the unit vectors $\vec{a}$, $\vec{b}$, $\vec{c}$ and $\vec{d}$ can lead to a
violation of Bell's inequality\cite{CHSH}:

\begin{equation}
E(\vec{a},\vec{b})+E(\vec{b},\vec{d})+E(\vec{c},\vec{d})-E(\vec{a},\vec{c})\le 2
\end{equation}

So the HF wave function is not entangled.

For a superposition of spin conserved Slater determinants, a spin adapted CI wave function, it 
can be proved that it violates Bell's inequality. For an $N$-electron system, the dimension of the configuration space is $D$, the CI
wave function is:

\begin{eqnarray}
|\Psi> & = & \sum_i^D c_i |\psi_i>  \nonumber \\
       & = & \sum_i^D c_i |\chi_{j}^{i}(1)\chi_{k}^{i}(2)...\chi_{l}^{i}(N)>
\end{eqnarray}

where $i$ represents $i$-th Slater determinant. We define the operator $\hat{O}$:

\begin{eqnarray}
\hat{O} & = & \vec{\sigma}(1) \vec{a} [{\bigoplus_i P_{j}^{i} (1)}]\otimes
\vec{\sigma}(2) \vec{b} [{\bigoplus_i P_{k}^i (2)}]+
\vec{\sigma}(1) \vec{a} [{\bigoplus_i P_{j}^i (1)}]\otimes
\vec{\sigma}(3) \vec{b} [{\bigoplus_i P_{k}^i (3)}]+...+ \nonumber \\
&   & \vec{\sigma}(1) \vec{a} [{\bigoplus_i P_{j}^i (1)}]\otimes
\vec{\sigma}(2) \vec{b} [{\bigoplus_i P_{k}^i (N)}] + \nonumber \\
&   & \vec{\sigma}(2) \vec{a} [{\bigoplus_i P_{j}^i (2)}]\otimes
\vec{\sigma}(1) \vec{b} [{\bigoplus_i P_{k}^i (1)}]+
\vec{\sigma}(2) \vec{a} [{\bigoplus_i P_{j}^i (2)}]\otimes
\vec{\sigma}(3) \vec{b} [{\bigoplus_i P_{k}^i (3)}]+...+ \nonumber \\
&   & \vec{\sigma}(2) \vec{a} [{\bigoplus_i P_{j}^i (2)}]\otimes
\vec{\sigma}(2) \vec{b} [{\bigoplus_i P_{k}^i (N)}] \nonumber \\
&   & +...+ \nonumber \\
&   & \vec{\sigma}(N) \vec{a} [{\bigoplus_i P_{j}^i (N)}]\otimes
\vec{\sigma}(1) \vec{b} [{\bigoplus_i P_{k}^i (1)}]+
\vec{\sigma}(N) \vec{a} [{\bigoplus_i P_{j}^i (N)}]\otimes
\vec{\sigma}(2) \vec{b} [{\bigoplus_i P_{k}^i (2)}]+...+ \nonumber \\
&   & \vec{\sigma}(N) \vec{a} [{\bigoplus_i P_{j}^i (N)}]\otimes
\vec{\sigma}(N-1) \vec{b} [{\bigoplus_i P_{k}^i (N-1)}]
\end{eqnarray}

We can calculate the expectation value $E(\vec{a},\vec{b})$:

\begin{equation}
E(\vec{a},\vec{b})=\sum_i^D c_i^2 <\chi_{j}^i |\vec{\sigma} \vec{a}|\chi_{j}^i><\chi_{k}^i |\vec{\sigma} \vec{b}|\chi_{k}^i>
\end{equation}

In general it cannot be factorized, the nonfactorization
of the mean value implies that we can choose four unit vectors $\vec{a}$, $\vec{b}$, $\vec{c}$, and $\vec{d}$
can lead to the violation of Bell's inequality.

We take the ground state HF wave function of H$_2$ molecule as an example to show that this 
wave function does not violate Bell's inequality.
The HF wave function of the ground state H$_2$ is the Slater determinant:

An entangled state violates Bell's inequality, for example, the famous Bell state
which is known as the spin singlet state:

\begin{eqnarray}
|\psi> & = & \frac{|01> - |10>}{\sqrt{2}}
\end{eqnarray}

where $0$ represents spin up state and $1$ represents spin down state. Suppose this entangled 
pair is hold by Alice and Bob respectively, Alice performs a measurement of spin along 
any real three-dimensional unit vector $\vec{a}$, 
that is, she measures the observable $\vec{\sigma}.\vec{a}=\sigma_x a_x+\sigma_y a_y+ \sigma_za_z$,
where $(\sigma_x,\sigma_y,\sigma_z)$ are the spin Pauli matrices.
 The measurement 
will give $+1$ or $-1$. Suppose Alice measures $+1$, then she can predict with certainty that
Bob will measure $-1$ on his qubit if he also measures spin along the $\vec{a}$ axis. Similarly if
Alice measures $-1$, she can predict with certainty that Bob will measure $+1$ on his qubit.  
Suppose Alice can measure the observable $\vec{\sigma}.\vec{a}$ and $\vec{\sigma}.\vec{d}$; while 
Bob can measure the observable $\vec{\sigma}.\vec{b}$ and $\vec{\sigma}.\vec{c}$. All these measurement
will give result of $+1$ or $-1$. Alice and Bob will measure the shared pair of particles at the
same time and choose randomly the observable to measure. Then we can calculate the expectation value 
$E(\vec{a} \vec{b}-\vec{a} \vec{c}+\vec{b} \vec{d}+\vec{c} \vec{d})$ 
and get the Bell's inequality\cite{CHSH}:

\begin{equation}
|E(\vec{a} \vec{b}-\vec{a} \vec{c}+\vec{b} \vec{d}+\vec{c} \vec{d})|=|E(\vec{a}, \vec{b})-E(\vec{a}, 
\vec{c})|+|E(\vec{b}, \vec{d})+E(\vec{c}, \vec{d})|\le 2
\end{equation}
where $E(\vec{a}, \vec{b})$ is the mean value of the product of the outcomes of two spin measurements
 along the direction $\vec{a}$ and $ \vec{b}$, this result is also known as the 
CHSH inequality\cite{CHSH}. However, for an entangled pair of particles like shown in 
Eq. (9), the Bell's inequality will be violated. For example, if they choose the following observables:

\begin{equation}
\vec{\sigma}\vec{a}=Z_1,   \vec{\sigma}\vec{d}=X_1,  \vec{\sigma}\vec{b}=\frac{-Z_2-X_2}{\sqrt{2}},  \vec{\sigma}\vec{c}=\frac{Z_2-X_2}{\sqrt{2}}
\end{equation}

One can find that $E(\vec{a} \vec{b}-\vec{a} \vec{c}+\vec{b} \vec{d}+\vec{c} \vec{d})=2\sqrt{2} > 2$, violates Bell's inequality.

Take the HF ground state wave function of H$_2$ molecule as an example to 
show that this 
wave function does not violate Bell's inequality.
The HF wave function of the ground state H$_2$ is the Slater determinant:

\begin{eqnarray}
|\Psi> & = & |\sigma_g(\uparrow\downarrow)> \nonumber\\
       & = & \frac{1}{\sqrt{2}}\left |\begin{array}{cc}
\sigma_g(1)\alpha(1) &\sigma_g^{'}(1)\beta(1)\\
\sigma_g(2)\alpha(2) &\sigma_g^{'}(2)\beta(2)
\end{array}\right |\\
\nonumber \\
& = & \frac{1}{\sqrt{2}}[\sigma_g(1)\alpha(1)\sigma_g^{'}(2)\beta(2)-\sigma_g(2)\alpha(2)
\sigma_g(2)\alpha(2)]
\end{eqnarray}

Here we use the restricted HF (RHF) wave function, where one spatial orbital
can contain two spin
opposite electrons, the spatial part of the two electrons are the same, 
but can be distinguished spatially,
so we use $\sigma_g$ and $\sigma_g^{'}$ to represent them. We define the operator:

\begin{equation}
\hat{O}=\vec{\sigma}(1). \vec{a} P(1)\otimes
\vec{\sigma}(2). \vec{b} P^{'}(2)+
\vec{\sigma}(1) .\vec{b} P^{'}(1)\otimes
\vec{\sigma}(2). \vec{a} P(2)
\end{equation}

where $P$ and $P'$ are the projection operators for spatially different orbitals. 
It can be shown that the 
mean value $E(\vec{a},\vec{b})$ of the product of the outcomes of two spin 
measurements along the directions 
$\vec{a}$ and $\vec{b}$ in two spatially different regions has the following
expression:

\begin{eqnarray}
E(\vec{a},\vec{b}) & = & <\Psi|\hat{O}|\Psi> \nonumber\\
                   & = & \frac{1}{2}[<\alpha(1)|\vec{\sigma}(1) \vec{a}|\alpha(1)>
<\beta(2)|\vec{\sigma}(2) \vec{b}|\beta(2)>+ \nonumber \\ {}
& & {} <\alpha(2)|\vec{\sigma}(2) \vec{a}|\alpha(2)>
<\beta(1)|\vec{\sigma}(1) \vec{b}|\beta(1)>]
\end{eqnarray}

because of the indistinguishable property of the electrons, the above 
formula can be written as:

\begin{equation}
<\Psi|\hat{O}|\Psi>=<\alpha|\vec{\sigma} \vec{a}|\alpha>
<\beta|\vec{\sigma} \vec{b}|\beta>
\end{equation}

in which it can be factorized as tensor product. The occurrence of a factorized 
product of two mean values 
implies that no choice of the unit vectors $\vec{a}$, $\vec{b}$, $\vec{c}$ 
and $\vec{d}$ can lead to a 
violation of Bell's inequality\cite{CHSH}.
So the HF wave function is not entangled. 
For a superposition of Slater determinants, a CI wave function, it 
can be proved that it violates Bell's inequality. Still using H$_2$ system as an example, 
in the STO-3G basis, when the distance $R$ between the two hydrogen atoms
goes to infinity, the state can be written as:

\begin{eqnarray}
\frac{1}{\sqrt{2}}(|\sigma_{g}(\uparrow\downarrow)>
+|\sigma_u(\uparrow\downarrow)>) & = & \frac{1}{2}[\sigma_{g}(1)
\alpha(1)\sigma_{g}^{'}(2) \beta(2)-
\sigma_{g}^{'} (1) \beta(1) \sigma_{g} (2) \alpha(2)+ {}
 \nonumber\\
 & & {}\sigma_{u}^{'}(1) \alpha(1) \sigma_{u}(2) \beta(2)-
\sigma_{u}(1) \beta(1) \sigma_{u}^{'}(2) \alpha(2)]
\end{eqnarray}

For the operator:
\begin{eqnarray}
\hat{O} & = & \vec{\sigma}(1). \vec{a} (P_{\sigma_g}(1) \oplus P_{\sigma_u}(1))\otimes
\vec{\sigma}(2). \vec{b} (P_{\sigma_{g}^{'}}(2) \oplus P_{\sigma_{u}^{'}}(2))+  {}
 \nonumber\\
 & & {}\vec{\sigma}(1) .\vec{b} (P_{\sigma_{g}^{'}}(1) \oplus P_{\sigma_{u}^{'}}(1))\otimes
\vec{\sigma}(2). \vec{a} (P_{\sigma_g}(2) \oplus P_{\sigma_u}(2))
\end{eqnarray}

It can be shown that the mean value $E(\vec{a}, \vec{b})$ is:

\begin{eqnarray}
E(\vec{a}, \vec{b}) & = & <\Psi|\hat{O}|\Psi> \nonumber \\
                    & = & \frac{1}{2}[<\alpha|\vec{\sigma} \vec{a}
|\alpha><\beta|\vec{\sigma} \vec{b}|\beta>+
 <\alpha|\vec{\sigma} \vec{b}|\alpha>
<\beta|\vec{\sigma} \vec{a}|\beta>]
\end{eqnarray}

in which the mean value can not be factorized as a tensor product, the non-factorization 
of the mean value implies that we can choose four unit vectors 
$\vec{a}$, $\vec{b}$, $\vec{c}$, and $\vec{d}$ 
can lead to the violation of Bell's inequality\cite{CHSH}: 

\begin{equation}
|E(\vec{a}, \vec{b})-E(\vec{a}, \vec{c})|+|E(\vec{b}, \vec{d})+E(\vec{c}, \vec{d})|\le 2
\end{equation}

So the CI wave function is entangled. It can be easily proved that any CI wave 
function is entangled.

\section{Measure the entanglement in systems of identical particles}
Shi\cite{shi1} suggested that entanglement 
of  identical particles can be characterized by using the 
antisymmetrized basis since they are not entangled,
which is equivalent to the particle number representation in the 
Fock space. This leads to the use
of occupation numbers of different single particle basis states as the
distinguishable degrees of freedom in quantifying entanglement of 
identical particles. The
non-factorization due to antisymmetrization is naturally excluded in this representation.
The occupation number entanglement in a system of fixed number of electrons is nothing
but the situation that the state is a superposition of different Slater determinants.

The particle number basis state for a fixed number of electrons
is the normalized antisymmetrized basis in configuration space, that is,
the Slater determinants. In terms of the product basis $|k_{1},...,k_{N}>$,
the $N$-particle state is

\begin{equation}
|\psi>=\sum_{k_1,...k_N}q(k_1,...,k_N)|k_1,...,k_N>
\end{equation}

the coefficients $q(k_1,...,k_N)$ are antisymmetric. $k_{i}$ is the single 
particle state.
For a fixed number of electrons, the normalized antisymmetrized basis can be 
written in terms of
the occupation numbers of different single-particle basis states. 
This is the particle number
representation, in which,

\begin{equation}
|\psi>=\sum_{n_1,...n_\infty}f(n_1,...,n_\infty)|n_1,...,n_\infty>
\end{equation}

where $n_j$ is the occupation number of mode $j$, which is the
spin orbital in molecules. The summation is subject to the constraint
$\sum_k n_k=N$, where $N$ is the total number of electrons.

A Slater determinant is not entangled with respect to
the given single-particle basis. A superposition of Slater determinants is
entangled in a given single-particle basis, that is, the spin orbital in chemical systems. 
A transformation from a superposition of Slater determinants
to a single Slater determinant in another single-particle
will involve operations on all particles in representing the state, it is a nonlocal
operation. Hence, the entanglement in the superposition of Slater determinants is the
entanglement between different single-particle basis states, the spin orbitals.

The density matrix of a chemical system can be obtained from
the wave function defined in the Fock space,
from which one obtains the single particle reduced density matrices
and the occupation numbers for each spin orbital. From
the Fock space reduced density matrices, the bipartite entanglement between
the occupation numbers of $l$ spin orbital and the occupation numbers
of the other spin orbitals can be calculated.

For electrons in the molecules, the single-particle basis is the
molecular spin orbital, which includes both spin and spatial orbital.
With the spatial orbital modes as labels with which the particles
are effectively distinguished, the entanglement can be viewed
as the spin entanglement among the particles in different spatial orbital modes.
In this way it is meaningful to say that the particle in a certain
orbital is spin entangled with the particle in another orbital and this spin
entanglement can be transformed into the spatial orbital mode entanglement.

As we have discussed before,
the entanglement in the superposition of different Slater determinants can be
expressed as the occupation number entanglement between spin orbitals
in a systems of fixed number of particles. Thus, entanglement is related to 
the electron correlation energy and is given by 
von Neumann entropy\cite{Peres}

\begin{equation}
S=-\sum_k \frac{n_k}{2} \log_2 \frac{n_k}{2} 
\end{equation}

where $n_k$ is the occupation number on the $k$-th spatial orbital.

\section{Entanglement and correlation energy calculations for the H$_2$ molecule}
The entanglement of superposition of Slater determinants finally results in 
the entanglement between the molecular spin orbitals, since different Slater determinants 
are constructed by different excitations between the spin orbitals. 
The spin orbital entanglement characterizes 
the entanglement in the superposition of Slater Determinants.
It can be characterized by using the 
von Neumann entropy of the occupation numbers of each spin orbital.
This occupation number is between $0$ and $1$.

The entanglement between the molecular spin orbitals 
is defined using the eigenvalues $n_k$
of the one-particle density matrix, which is the occupation of the spin orbitals.  
Since we use the restricted Hartree-Fock wave function in the calculation, two 
electrons share the same spatial orbital and only different in the spin part, the 
occupation of each spin orbital is one-half of the occupation of the spatial orbital,
the occupation number is between $0$ and $1$. So the entanglement 
between the spin orbitals in the system can be calculated using Eq. (22),
which characterizes the entanglement between these $N/2$ spin orbitals and 
the other $N/2$ spin orbitals,  
where $n_k$ is the occupation number on the $k$-th spatial orbital. In fact, similar 
formula to Eq.(21) have been used 
in the study of the entanglement in Hubbard model\cite{zie1} and correlation 
entropy of the H$_2$ molecule\cite{zie2}. In this paper we emphasized the physical meaning
of the formula and show that it measures the spin entanglement of the system.

Electron correlation effects in the molecular system end up as a 
function of the one particle
density matrix, i.e. the occupation numbers of each orbital, thus we can see
that through one particle density matrix, entanglement and electron correlation can
be connected. We studied the entanglement and electron correlation energy in
hydrogen molecule, using two basis sets: the STO-3G and 6-31G** 
 to study the relation between these two quantities.
As shown in the figures both results support our expectation that electron correlation
is related quantum entanglement in chemical system.

The  entanglement for the H$_2$ molecule as a function of $R$, the 
distance between two hydrogen atoms, can be evaluated 
using a one-particle density matrix calculated from
the FCI wave function. Fig. 1 shows the calculation using STO-3G basis, 
Fig. 2 using the 6-31G** basis set.  
The correlation energy is 
calculated as $E_{corr}=E_{HF}-E_{FCI}$.
Both the entanglement and the correlation energy goes to $0$ as 
$R \rightarrow 0$ in the STO-3G calculation. 
As $R$ increases, the entanglement and the correlation energy increase, 
when $R \rightarrow \infty$,
the entanglement reaches its maximum $1$, and the correlation energy 
also reaches its maximum. 
This can be explained that as $R \rightarrow \infty$, the overlap of the 
two hydrogen atomic wave functions goes to $0$, the energy difference 
between the two molecular 
orbitals in the H$_2$ system, $\sigma_g$ and $\sigma_u$ goes to $0$, the 
contribution to the 
configuration interaction from the $\sigma_u$ orbital is
equivalent to that of the $\sigma_g$ orbital, so their occupations are equal,
which makes the entanglement of the system goes to the maximum $1$. So as 
$R \rightarrow \infty$,
the system is in a superposition state,

\begin{equation}
\frac{1}{\sqrt{2}}(|\sigma_g(\uparrow)\sigma_g(\downarrow)>
+|\sigma_u(\downarrow)\sigma_u(\uparrow)>)
\end{equation}

which is entangled. This will be clear when it is written in the expanded form:

\begin{eqnarray}
\frac{1}{\sqrt{2}}(|\sigma_{g}(\uparrow\downarrow)>
+|\sigma_u(\uparrow\downarrow)>) & = & \frac{1}{2}[\sigma_{g}(1)
\alpha(1)\sigma_{g}(2) \beta(2)-
\sigma_{g} (1) \beta(1) \sigma_{g} (2) \alpha(2)+ \nonumber\\
&   & \sigma_{u}(1) \alpha(1) \sigma_{u}(2) \beta(2)-
\sigma_{u}(1) \beta(1) \sigma_{u}(2) \alpha(2)] \nonumber \\
& = & \frac{1}{2}(\sigma_{g}(1)\sigma_{g}(2)+\sigma_{u}(1) \sigma_{u}(2))
 (\alpha(1) \beta(2)-
 \beta(1) \alpha(2))                                
\end{eqnarray}

 As $R$ decreases, the entanglement and the correlation energy decrease, 
the system is in the state:
$\alpha |\sigma_g(\uparrow)\sigma_g(\downarrow)>+
\beta |\sigma_u(\downarrow)\sigma_u(\uparrow)>$
where $\alpha^2 + \beta^2 = 1$ and $\alpha \ne \beta$.
When $\alpha \ne \beta$, the entanglement is less than $1$.
When $R \rightarrow 0$,
the entanglement reaches its minimum $0$, and the correlation energy also reaches $0$.
This can be explained that as $R \rightarrow 0$, the overlap of the
two hydrogen atomic wave function increases and goes to $1$ when $R=0$, 
that is the He atom.
The energy difference between the two molecular
orbitals in the H$_2$ system, $\sigma_g$ and $\sigma_u$ increases and goes to $\infty$ 
when $R$ goes to $0$, the contribution to the
configuration interaction from the $\sigma_u$ orbital goes to $0$,
so the occupation on $\sigma_u$ orbital decreases to $0$ and the occupation 
on $\sigma_g$ orbital increases to $2$. 
As $R=0$, for the He atom the two electrons are located in the $1s$ orbital, 
the system is completely described by one Slater determinant, the entanglement of the 
system goes to $0$ and there is no correlation energy. 
We can see from the figure that the 
correlation energy changes in the same way as the entanglement changes, 
this supports our argument 
at the   perfectly. We also studied the system using a bigger basis 
set the 6-31G** which gives us similar results, except that there are 
occupations on the other molecular orbitals and both the entanglement and 
the electron correlation energy are not $0$ when $R \rightarrow 0$.  
We can also see that as the basis set becomes larger,
the rescaled entanglement and the correlation energy become closer.      
The rescaled entanglement is simply  obtained by
 multiplying it by a constant  to force the entanglement at infinity to 
be equal to the electron correlation energy.

\section{Exchange correlation and entanglement}

Next we will discuss the relation between spin exchange correlation and 
quantum entanglement. 
As $R \rightarrow \infty$, the electron correlation energy is 
equal to the exchange correlation energy,
which is the exchange integral that quantifies spin 
correlation in a molecular system.  
In the STO-3G calculation, the electron correlation 
energy is given as\cite{szabo}:

\begin{equation}
 E_{corr}=\Delta-\sqrt{\Delta^2 + K^2_{12}}
\end{equation}

where,

\begin{equation}
 \Delta = \frac{1}{2}(2(\varepsilon_2 - \varepsilon_1) + J_{11} + 
J_{22} - 4 J_{12} + 2 K_{12})
\end{equation}

where $\varepsilon_1$ and $\varepsilon_2$ are the energies of the $\sigma_g$ and 
$\sigma_u$ orbitals respectively, 
$J_{11}$, $J_{22}$ and $J_{12}$ are the Coulomb integrals for Coulomb 
repulsion between electrons 
on $\sigma_g$ orbital, $\sigma_u$ orbital and between $\sigma_g$ orbital and 
$\sigma_u$ orbital, 
$K_{12}$ is the exchange integral. 

When $R \rightarrow \infty$, the energy difference 
between the $\sigma_g$ and $\sigma_u$ orbitals goes to $0$, the $\sigma_g$ and  
$\sigma_u$ orbitals are degenerate. For these two degenerate and orthogonal orbitals,  
$K_{12}$ reaches its maximum as $R \rightarrow \infty$, meanwhile
$J_{11}$, $J_{22}$ and $J_{12}$ reach their minimum.
The Coulomb integral $J_{11}$, $J_{22}$, $J_{12}$ 
and $K_{12}$ equal to each other. Such that when $R \rightarrow \infty$, 
the correlation energy is equal to the exchange integral, 
$E_{corr} = -K_{12}$. 
So as $R \rightarrow \infty$, the exchange
integral measures the spin correlation and can be used to characterize 
the spin entanglement, which 
is related to the occupation number entanglement, which reaches the maximum 
as the $\sigma_g$ and $\sigma_u$ orbitals are degenerate. We expect the same
behavior upon increasing the basis set to 6-31G**.

\section{Summary}

The concept of electron correlation as 
defined in quantum chemistry calculations
is useful but not directly observable,
i.e., there is no operator
in quantum mechanics that its measurement gives the correlation energy. 
Moreover, there are cases where the 
kinetic energy dominates the Coulomb repulsion between 
electrons, the  electron correlation alone fails as a 
correlation measure\cite{ziescheJCP,zhen-CPL}. 
Here, we  have studied the relation between the electron correlation energy and 
the entanglement in H$_2$ system.
We show that entanglement can be used as an alternative measure of 
electron correlation
in quantum chemistry calculations. The spin entanglement can be measured using the 
occupation number entanglement between spin orbitals.  Work is underway to investigate
entanglement for larger molecular systems.

\newpage

\begin{figure}
\begin{center}
\includegraphics[width=0.9\textwidth,height=0.6\textheight]{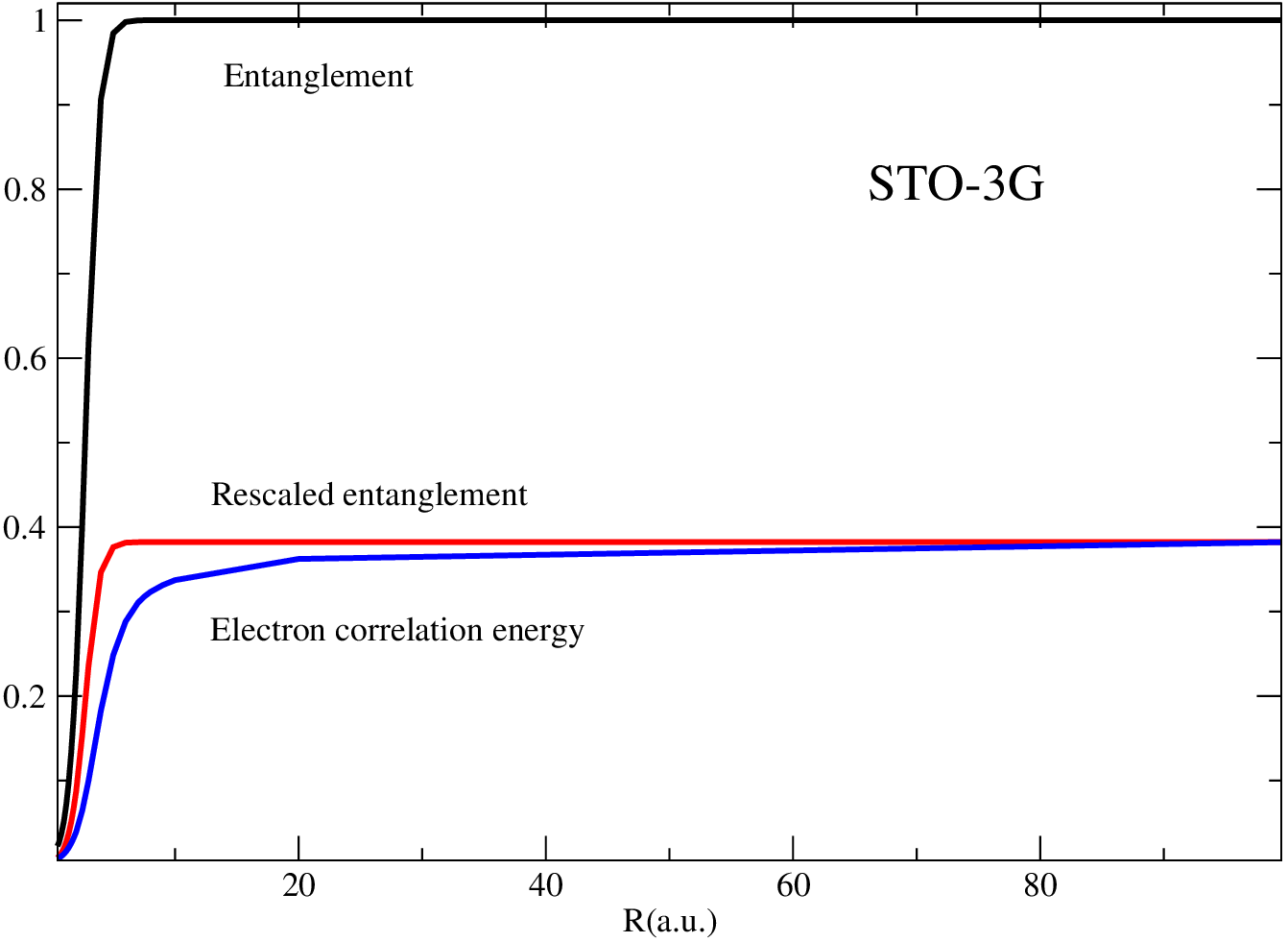}
\end{center}
\caption{Comparison between entanglement and electron correlation 
energy as a function of the internuclear distance $R$ for the 
 H$_2$ molecule using the 
Gaussian basis set  STO-3G.}  
\end{figure}

\newpage

\begin{figure}
\begin{center}
\includegraphics[width=0.9\textwidth,height=0.6\textheight]{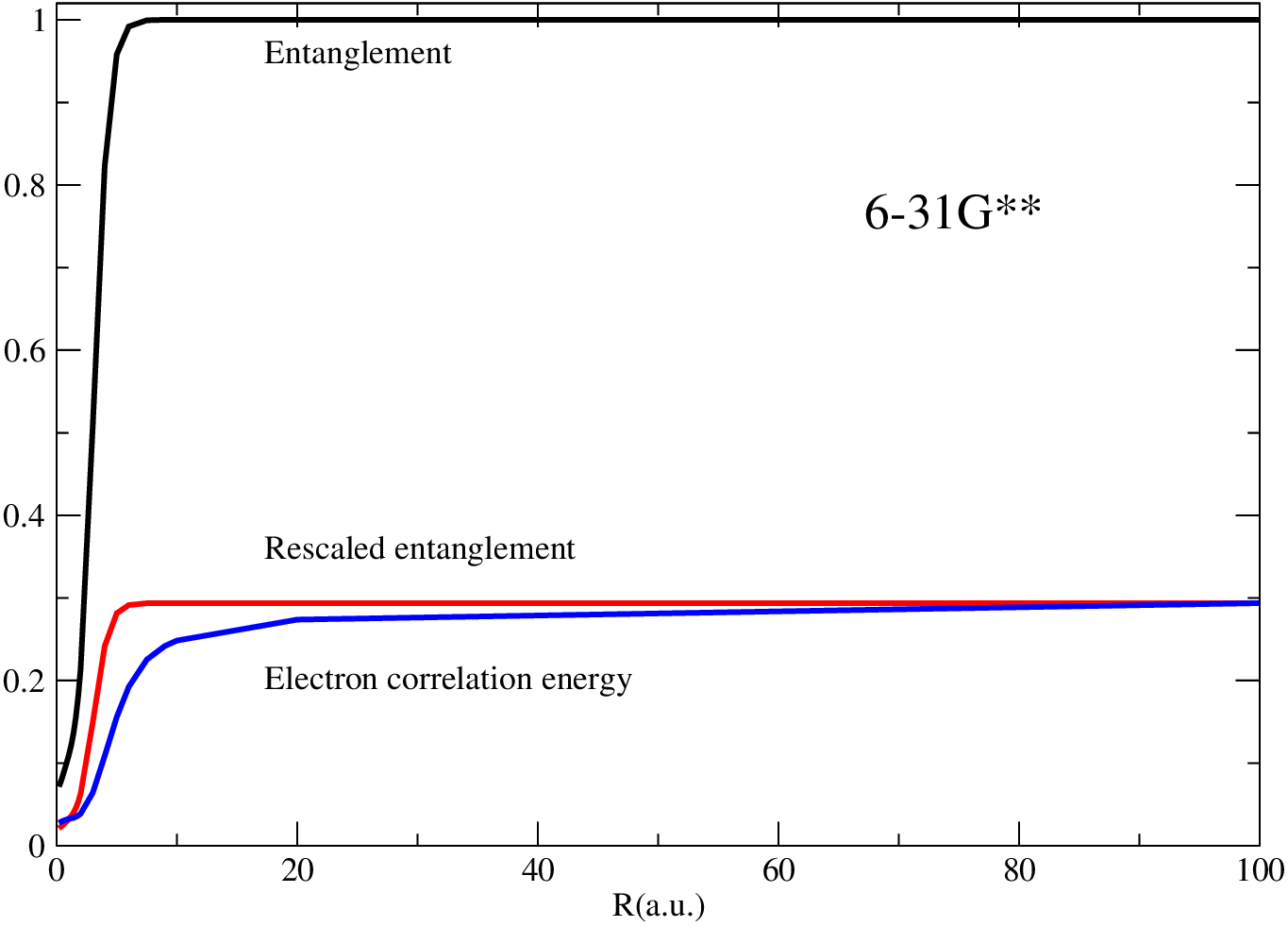}
\end{center}
\caption{Comparison between entanglement and electron correlation 
energy as a function of the internuclear distance $R$ for the 
 H$_2$ molecule using  the Gaussian basis set 6-31G**.}
\end{figure}

\end{document}